%% file: PR-UoS.tex
\documentclass{article}
\pdfoutput=1 
\usepackage{spconf,amsmath,graphicx,enumitem}

\usepackage[font=footnotesize,labelfont=bf]{caption}
\usepackage{color,mathtools,amssymb}
\usepackage{balance}

\graphicspath{{figs/}}

\input{latexdefs.tex}

\input{definitions}
\title{Phase Retrieval for Signals in Union of Subspaces}
\twoauthors
  {M. Salman Asif\sthanks{Email: sasif@ucr.edu. This work is supported in part by UCR Regents Faculty Fellowship. }
  }
	{Electrical and Computer Engineering\\
	University of California, Riverside}
  {Chinmay Hegde\sthanks{Email: chinmay@iastate.edu. This work is supported in part by grants CCF-1750920 and CCF-1815101 from the National Science Foundation, and a faculty fellowship from the Black and Veatch Foundation.}}
	{Electrical and Computer Engineering\\
	Iowa State University}

\begin{document}
%
\maketitle
\begin{abstract}
We consider the phase retrieval problem for signals that belong to a union of subspaces. We assume that amplitude measurements of the signal of length $n$ are observed after passing it through a random $m \times n$ measurement matrix. We also assume that the signal belongs to the span of a single $d$-dimensional subspace out of $R$ subspaces, where $d\ll n$. We assume the knowledge of all possible subspaces, but the true subspace of the signal is unknown. We present an algorithm that jointly estimates the phase of the measurements and the subspace support of the signal. We discuss theoretical guarantees on the recovery of signals and present simulation results to demonstrate the empirical performance of our proposed algorithm. Our main result suggests that if properly initialized, then $O(d+\log R)$ random measurements are sufficient for phase retrieval if the unknown signal belongs to the union of $R$ low-dimensional subspaces.

\end{abstract}
\begin{keywords}
alternating minimization, subspace identification, block sparsity.  
\end{keywords}

\section{Introduction}
\input{intro}
\section{Background} 
\input{background}

\section{Algorithm} 
\input{algo}

\section{Analysis}
\input{analysis}

\section{Simulation Results} 
\input{simulations}


\ninept 

\balance
\bibliographystyle{IEEEbib}
\bibliography{PR-UoS,biblio_isit,biblio_nips,chinbiblio}

\end{document}

%% file: latexdefs.tex
\usepackage{graphicx}
\usepackage{amsmath,amsfonts,amsthm,amssymb,xspace,bm, verbatim,dsfont}
\usepackage{url,algorithm,algpseudocode}
\usepackage{thmtools}
\usepackage{thm-restate}
\usepackage{cleveref}
\usepackage{tikz}
\usetikzlibrary{decorations.pathreplacing,angles,quotes}
\usetikzlibrary{positioning,plotmarks,calc}
\usepackage{bbm}
\usepackage{todonotes}
\usepackage{subfig}
\usepackage{pgfplots}
  
\theoremstyle{plain}

\theoremstyle{definition}
\newcommand{\note}[1]{\marginpar{\tiny *note in TeX*}}

\newcommand{\reals}{\mathbb{R}}

\newcommand{\vect}[1]{\mathbf{#1}}
\newcommand{\mat}[1]{\mathbf{#1}}

\newcommand{\iprod}[2]{\left\langle #1, #2 \right\rangle} 
\newcommand{\abs}[1]{\left|#1\right|}

\newcommand{\twonorm}[1]{\left\| {#1} \right\|_2}

\newcommand{\sign}[1]{\operatorname{sign}\left(#1\right)}

\newcommand{\distop}[2]{\mathrm{dist}\left(#1,#2\right)}

\DeclareMathOperator*{\argmin}{argmin}
\DeclareMathOperator*{\argmax}{argmax}

\newcommand{\rbrak}[1]{\left(#1\right)}

\newcommand{\cbrak}[1]{\left\{#1\right\}}

\newcommand{\y}{\vect{y}}
\newcommand{\e}{\vect{e}}
\newcommand{\z}{\vect{z}}

\newcommand{\xo}{\vect{x^*}}
\newcommand{\xin}{\vect{x^0}}
\newcommand{\x}{\vect{x}}
\newcommand{\xt}{\x^t}
\newcommand{\xtplus}{\x^{t+1}}

\newcommand{\ai}{\vect{a}_i}

\newcommand{\p}{\vect{p}}

\newcommand{\A}{\mat{A}}

\newcommand{\U}{\mat{U}}
\newcommand{\subspaces}{{\mathcal{M}}}

%% file: definitions.tex
\newcommand{\mA}{\ensuremath{\mathbf A}}

\newcommand{\mG}{\ensuremath{\mathbf G}}

\newcommand{\mQ}{\ensuremath{\mathbf Q}}

\newcommand{\va}{\ensuremath{\mathbf a}}

\newcommand{\vx}{\ensuremath{\mathbf x}}
\newcommand{\vy}{\ensuremath{\mathbf y}}

%% file: intro.tex
Phase retrieval refers to a broad class of problems in which we seek to recover a real- or complex-valued signal from its amplitude measurements. 
We assume that an unknown signal, $\xo \in  \reals^n$, is measured via (possibly noisy) observations of the form:
\begin{align}
\label{eqn:magnitude-measurements}
y_i = \abs{\iprod{\ai}{\xo}} + e_i,\hspace{0.5cm} i=1,\ldots,m,  
\end{align}
Throughout this paper, we assume that the measurements are \emph{Gaussian}, i.e., each element of $\ai \sim \mathcal{N}(0,1)$. The task is to recover an estimate of $\xo$ from the (phaseless) measurements $\y \in \reals^m$. 

\begin{figure}[!h]
	\centering
	\includegraphics[width=0.3\linewidth]{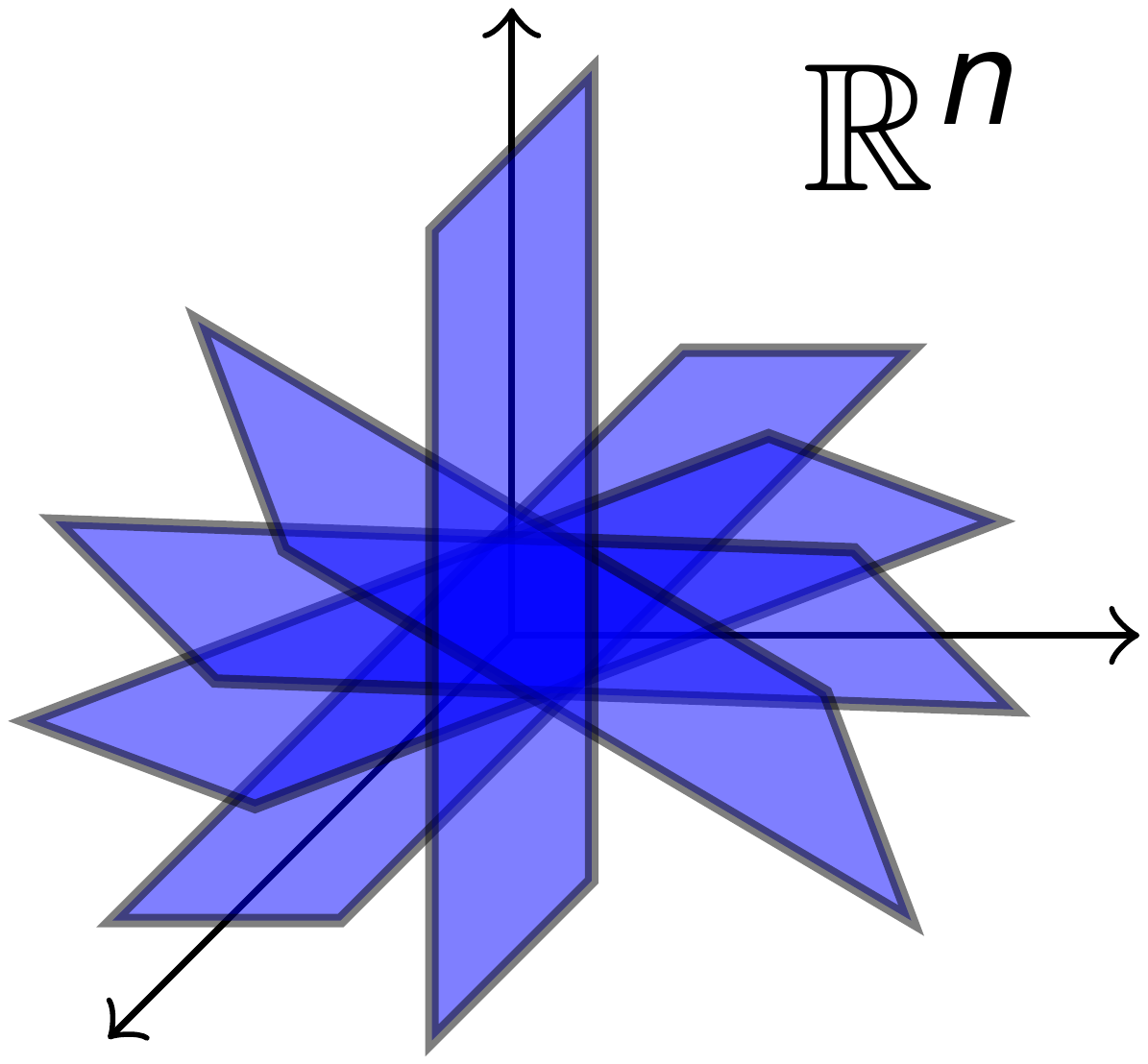}
	\caption{Illustration of a union-of-subspaces.}\label{fig:UoS}
\end{figure}

Our focus is on the special case where $\xo$ is assumed to belong to a \emph{union of subspaces}. Formally, let us assume that we are given a collection of we represent the union of subspaces as 
\[
\mathcal{M} = \{ \x \in \reals^n | \x = \U_i \alpha ~~ \text{for some }i\in\{1,\ldots,R\}\},
\]
where $\U_i \in \{\U_1, \U_2, \ldots, \U_R \}$ and each $\U_i \in \reals^{n \times d}$ is an orthonormal basis for a $d$-dimensional subspace of $\reals^n$ and $\alpha \in \reals^d$ is a coefficient vector. Figure~\ref{fig:UoS} is a depiction of a union of subspaces (planes). 
Throughout this paper, we assume that $\subspaces$ is known, and that $\xo \in \subspaces$ (i.e., we consider no model mismatch). For the rest of this paper, we ignore noise (i.e., we set $e_i = 0$) while noting that most of our arguments carry over under sub-gaussian assumptions on the noise variables $e_i$. 

We present a two-stage algorithm in which we first construct a coarse initial estimate of the signal using spectral initialization. Then we iteratively refine the signal estimate using an alternating minimization method in which we alternately solve a least-squares problem with union-of-subspace constraints and update the measurement phase. 

We analyze the performance of our proposed algorithm with theoretical guarantees. Our main result is given in Theorem~\ref{thm:lin_convergence}, which suggests that $m > C(d+\log R)$ i.i.d. Gaussian measurements are sufficient for the recovery of any signal $\vx^* \in \subspaces$, where $C>0$ is a constant factor. We also analyze the empirical performance of our algorithm using simulated measurements. The results are summarized in Figure~\ref{fig:results}. 

%% file: background.tex
The phase retrieval problem has been extensively studied over the last few decades \cite{gerchberg1972phase,fienup1982phase,candes2013phaselift} and it appears in several applications, including optical imaging \cite{fienup1982phase,holloway2016toward}, microscopy \cite{tian2014multiplexed,rodenburg2008ptychography}, and X-ray crystallography \cite{miao1999extending}. 

Phase retrieval is a non-convex problem and classical solution methods rely on alternating projection heuristics; examples include Gerchberg-Saxton  \cite{gerchberg1972phase} and Fienup algorithms \cite{fienup1982phase}. In recent years, lifting-based methods were introduced that reformulate the phase retrieval as a rank-one matrix recovery problem and relax it to a nuclear norm minimization problem that can be solved as a semidefinite program \cite{candes2013phaselift}. Because of large computational complexity and memory requirements of semidefinite program-based approaches, they are infeasible for large-scale problems. More recently, a number of convex and non-convex methods have been proposed for solving phase retrieval problem with theoretical performance guarantees  \cite{bahmani,goldstein2016phasemax,netrapalli,wang2017solving,candes2015codeddiff,jagatap2017fast}. Almost all these methods rely on estimating a good initial solution via the so-called \textit{spectral initialization} method. In its simplest form, a spectral initializer computes the top singular vector of the following Hermitian matrix: 
\begin{equation}
\mG = \frac{1}{m}\sum_{i=1}^m y_i^2 \va_i\va_i^T,
\end{equation}
which is equivalent to finding a unit norm vector $\vx$ that maximizes the inner product between the observed amplitude vector $\vy$ and $|\mA\vx|$. In our proposed algorithm, we use a similar approach for initialization in which we estimate an initial solution in every subspace and select the one that maximize the inner product. 


%% file: algo.tex
Our algorithm follows a two-stage approach that has become common in the phase retrieval literature. 
\begin{itemize}
\item \emph{Initialization}: We first construct a coarse initial estimate of the signal. Specifically, we obtain a signal estimate $\x^0$ using spectral initialization process described in Algorithm~\ref{alg:init}. 
Such an estimate is expected to satisfy the following condition: 
\[
\distop{\x^0}{\xo} \leq \delta \twonorm{\xo} 
\]
for some small constant $\delta$, where $\distop{\cdot}{\cdot}$ is a suitably defined distance measure.

\item \emph{Refinement}: We then sequentially refine this estimate using a strategy based on \emph{alternating minimization}, following a variant of~\cite{netrapalli,copram,copramext}. We prove that our refinement strategy demonstrates linear convergence to $\xo$, i.e., for $t = 0, 1, 2, \ldots$, our sequence of estimates satisfies:
\[
\distop{\x^{t+1}}{\xo} \leq \rho \, \distop{\x^{t}}{\xo}.
\]
\end{itemize}

%% file: analysis.tex
Our overall algorithm is described in pseudocode form in Algorithm~\ref{alg:phase_subspace}. In this section, we discuss  some theoretical guarantees for subspace identification and signal reconstruction using our proposed two-step method. 

\renewcommand{\algorithmicrequire}{\textbf{Input:}}
\renewcommand{\algorithmicensure}{\textbf{Output:}}
\begin{algorithm}[!t]
	\caption{PR-UoS: Spectral initialization}
	\label{alg:init}
	\begin{algorithmic}[1]
		\Require $\A,\y, \{\U_1, \ldots, \U_R\}$
		\item[]
		\State	Compute top singular vector for every subspace as 
		\begin{equation}\label{eq:init}
		\hat{\alpha}_r = \argmax_{\alpha: \|\alpha\|_2 = 1}\;\|\y\circ \A\U_r\alpha\|_2.
		\end{equation}
		\State	Select the subspace that maximizes the objective in \eqref{eq:init}: $$r^* = \argmax_r\;\|\y\circ \A\U_r\hat\alpha_r\|_2.$$ %
		\item[]
		\Ensure $\x^{0} \leftarrow \U_{r^*}\alpha_{r^*}$.
	\end{algorithmic}
\end{algorithm}
\begin{algorithm}[!t]
	\caption{PR-UoS: Descent}
	\label{alg:phase_subspace}
	\begin{algorithmic}[1]
		\Require $\A,\y, \subspaces$
		\item[]
		\State Initialize $\x^0$ according to Algorithm~\ref{alg:init}
		\For{$t = 1,\cdots,T$}
		\State	$\vect{p}^{t} \leftarrow \sign{\A \vect{x}^{t-1}}$,
		\State	$\vect{x}^{t} = \argmin_{\x \in \subspaces} \twonorm{\vect{p}^{t} \circ \y - \A \x}$
		\EndFor
		\item[]
		\Ensure $\hat \x \leftarrow \vect{x}^T$.
	\end{algorithmic}
\end{algorithm}

\subsection{Analysis of Descent} 
\label{subsec:descent}
This part of the algorithm is described in Lines 2-5 of Algorithm~\ref{alg:phase_subspace}. Once we obtain a good enough initial estimate $\xin \in \mathcal{M}$ such that $\distop{\xin}{\xo} \leq \delta_0 \twonorm{\xo}$, we construct a method to refine this estimate. 

The intuition is as follows. The observation model in \eqref{eqn:magnitude-measurements} can be restated as follows:
\begin{align*}
\sign{\iprod{\ai}{\xo}}\circ y_i = \iprod{\ai}{\xo} ,
\end{align*}
for all $i=\{1,2,\ldots, m\}$. To ease notation, denote the \emph{phase vector} $\vect{p} \in \reals^m$ as a vector that contains the unknown signs of the measurements, i.e., ${p}_i = \sign{\iprod{\ai}{\x}}$ for all $i=\{1,2,\ldots,m\}$. Let $\vect{p}^*$ denote the true phase vector and let $\mathcal{P}$ denote the set of all phase vectors, i.e. $\mathcal{P} = \cbrak{\vect{p}:p_i = \pm 1, \forall i}$. Then our measurement model gets modified as:
\begin{align*}
\vect{p}^*\circ\y = \A \xo.
\end{align*}

Therefore, the recovery of $\xo$ can be posed as a (non-convex) optimization problem:
\begin{align} \label{eq:lossfunc}
\min_{\x \in \mathcal{M},\vect{p} \in \mathcal{P}} \twonorm{\A\x - \vect{p}\circ\y}
\end{align} 

To solve this problem, we alternate between estimating $\vect{p}$ and $\x$.
We perform two estimation steps: 

\begin{enumerate}
\item if we fix the signal estimate $\x$, then the minimizer $\vect{p} \in \mathcal{P}$ is given in closed form as:
	\begin{align} \label{eq:phase_est}
	\vect{p}=\sign{\A\x} ,
	\end{align} 
\item and if we fix the phase vector $\vect{p}$, the signal vector $\x \in \mathcal{M}_s$ can be obtained by solving a sparse recovery problem,
	\begin{align} \label{eq:loss_min}
	\min_{\x \in \mathcal{M}} \|\A \x-\vect{p}\circ \y\|_2 .
	\end{align}
\end{enumerate}

We now analyze our proposed descent scheme. We obtain the following theoretical result:
\begin{restatable}{theorem}{convergence}
\label{thm:lin_convergence}
	Given an initialization $\xin \in \mathcal{M}$ satisfying $\distop{\x^0}{\xo} \leq \delta \twonorm{\xo}$, for $0 < \delta_0 < 1$, if the number of (Gaussian) measurements,
	$$ m >  C \rbrak{d + \log R}, $$ 
	then with high probability, the iterates $\x^{t+1}$ of Algorithm~\ref{alg:phase_subspace}, satisfy:
	\begin{align} \label{eq:mainconvergence}
	\distop{\xtplus}{\xo} \leq {\rho}\, \distop{\xt}{\xo},
	\end{align}
	where $\x^t,\x^{t+1}, \xo \in \mathcal{M}$, and $ 0 < \rho < 1$ is a constant.
\end{restatable}

\noindent{\textbf{Proof sketch:}} The high level idea behind the proof is that with a $\delta$-ball around the true signal $\xo$, the ``phase noise'' can be suitably bounded in terms of a constant times the signal estimation error. To be more precise, suppose that $\z^* = \A \xo = \p^*\circ\y$. Then, at any iteration $t$, we have:
\begin{align*}
\z^t &= \p^t \circ \y \\
&= (\p^t - \p^*) \circ \y +p^* \circ \y \\
&= \z^* + \e^t,
\end{align*}
where $\e^t$ can be viewed as the ``phase noise''. Now, examining Line 4 of the above algorithm, we have:
\begin{align*}
\twonorm{\A \x^t - \p^t \circ \y} &\leq \twonorm{\A \xo - \p^t \circ \y}~~(\text{feasibility of}~\xo) \\
&=\twonorm{\z^* - \z^t} \\
&= \twonorm{\e^t} .
\end{align*}
On the other hand, we have:
\begin{align*}
\twonorm{\A \x^t - \p^t \circ \y} &= \twonorm{\A \x^t - \z^t} \\
&\geq \twonorm{\A \x^t - \z^*} - \twonorm{\z^* - \z^t}\\
&= \twonorm{\A(\x^t - \xo)} - \twonorm{\e^t} \\
&\geq (1 - \delta_0) \twonorm{\x^t - \xo} - \twonorm{\e^t},
\end{align*}
where the second inequality follows from the triangle inequality and the last inequality follows from the restricted isometry property of $\A$ (with constant $\delta_0$) over the union of subspaces $\mathcal{M}$; for Gaussian $\A$, this holds provided:
\[
M \geq \frac{C}{\delta_0^2} (d + \log R) ,
\]
for any finite union of subspaces $\mathcal{M}$ \cite{blumensath2009iterative}. Rearranging the inequalities provide us the following bound: 
\[
\twonorm{\x^t - \xo} \leq \frac{2}{1 - \delta_0} \twonorm{\e^t} .
\]
It remains to show that $\twonorm{e^t}$ can be bounded in terms of $\twonorm{\x^{t-1} - \xo}$. We do this through Lemma \ref{lem:phase_err_bound} below. Consequently, we get 
\[
\twonorm{\x^t - \xo} \leq \frac{2 \rho'}{1 - \delta_0} \twonorm{\x^{t-1} - \xo},
\]
where $\rho'$ is a small enough constant.

We therefore achieve a per-step error reduction scheme of the form:
\begin{align*}
\twonorm{\x^{t+1} - \x^*} \leq \rho_0 \twonorm{\x^{t} - \x^*},
\end{align*}
if the initial estimate $\xin$ satisfies $\twonorm{\xin - \x^*} \leq \delta_0 \twonorm{\x^*}$, and this result can be trivially extended to the case where the initial estimate $\xin$ satisfies $\twonorm{\xin + \x^*} \leq \delta_0 \twonorm{\x^*}$, hence giving the convergence criterion of the form (for $\rho < 1$): 
\begin{align*}
\distop{\x^{t+1}}{\xo} \leq \rho\, \distop{\x^{t}}{\xo}.
\end{align*}

We now state Lemma \ref{lem:phase_err_bound} without proof. A proof will be provided in an extended version of this paper. The proof follows an adaptation of the approach in~\cite{copram,copramext}, which itself is based on the approach of \cite{zhang2016reshaped}.
\begin{restatable}{lemma}{phaseerror} \label{lem:phase_err_bound}
	As long as the initial estimate is a small distance away from the true signal $\xo \in \mathcal{M}$, $	\distop{\xin}{\xo} \leq \delta \twonorm{\xo}$,
	and subsequently,
	$\distop{\xt}{\xo} \leq \delta \twonorm{\xo}$, where $\x^t$ is the $t^{th}$ update of Algorithm~\ref{alg:phase_subspace}, then the following bound holds for any $t \geq 0$:
	\begin{align*}
	\twonorm{\e^{t+1}} \leq  \rho_1 \twonorm{\xt - \xo},
	\end{align*}
	with high probability, as long as $ m > C (d + \log R)$ and $\rho_1 < 1$ is a constant.
\end{restatable}

%% file: simulations.tex
In this section, we present some simulation results to demonstrate the performance of our proposed algorithm for different values of signal length $(n)$, number of subspaces $(R)$, and the dimension of each subspace $(d)$. The number of measurements $(m)$ in all our simulations is same as $n$. 
Figure~\ref{fig:results} summarizes our simulation results, where each point represents the probability of success in signal recovery for the given values of $n,d,R$. 

\newcommand{\figwidth}{0.31\textwidth}
\begin{figure*}[!ht]
	\centering 
	\includegraphics[width=\figwidth,keepaspectratio]{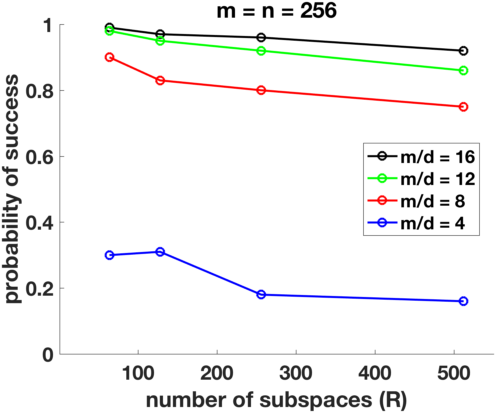}~~
	\includegraphics[width=\figwidth,keepaspectratio]{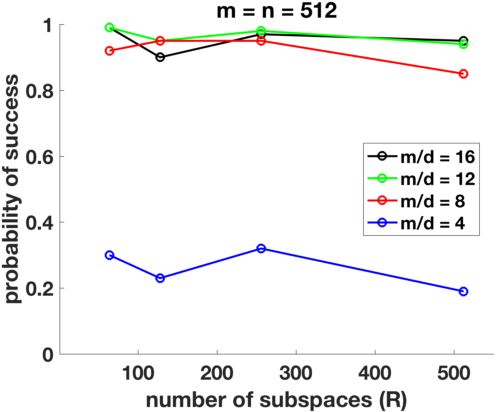}~~
	\includegraphics[width=\figwidth,keepaspectratio]{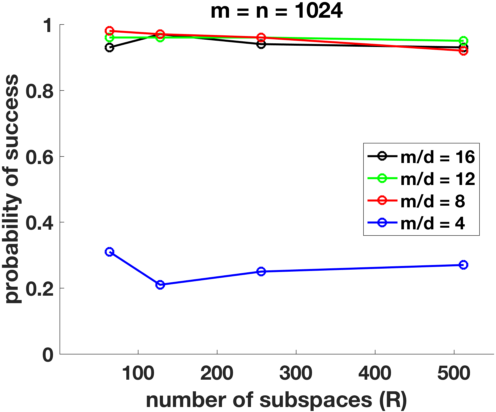}
	\caption{Probability of success in signal recovery for different values of $n, d, R$. $m = n$ in all the simulations.  }\label{fig:results}
\end{figure*}

Our simulation setup is as follows. For given values of $n,R,d$, we first generated an $n\times n$ random matrix $\mA$ with i.i.d. Gaussian entries. Then we selected an $n\times n$ matrix $\mQ$ with orthogonal columns and generated $\{\U_1, \ldots \U_R\}$ by selecting $d$ columns from $\mQ$ at random for each $\U_i$. We generated $\vx^* = \U_i \alpha$ by picking a $\U_i$ at random and creating $\alpha$ with i.i.d. Gaussian entries. We simulated measurements as $\vy = |\mA\vx^*|$ and estimated $\hat \vx$ using the two-stage phase retrieval process described in Algorithm~\ref{alg:phase_subspace}. We performed 100 independent trial for each tuple $n,R,d$. In every trial, we considered the recovery successful if the normalized difference between the true and estimated signals $$\frac{\|\vx^*-\hat \vx\|_2}{\|\vx^*\|_2} < \tau,$$ where $\tau = 10^{-5}$ is a threshold. 

We observe that as $m/d$ increases, the probability of success approaches 1. The reconstruction is almost always possible if $\frac{m}{d} > 8$. We also observe that increasing the number of subspaces $(R)$ has very little effect on the recovery performance once $\frac{m}{d}$ increases beyond a certain value.